\documentclass[12pt]{article}
\usepackage{latexsym}
\usepackage{amsmath}
\usepackage{bm}
\usepackage[dvipdfmx]{graphicx,color}
\usepackage{amssymb}
\usepackage{cite}
\usepackage{amsfonts, amscd, amsthm}
\usepackage[all]{xy}
\usepackage{here}
\usepackage{ulem}

\renewcommand{\theequation}{\thesection.\arabic{equation}}

\makeatletter
\@addtoreset{equation}{section}
\renewcommand{\theequation}{\thesection.\@arabic\c@equation}
\makeatother

\makeatletter
\renewcommand\appendix{\par
  \setcounter{section}{0}%
  \setcounter{subsection}{0}%
  \gdef\thesection{Appendix \@Alph\c@section }
  \renewcommand{\theequation}
  {\Alph{section}.\arabic{equation}}
}
\makeatother

\makeatletter
\newcounter{subeqncnt}
\def\thesubeqncnt{\alph{subeqncnt}}
\def\subequations{\begingroup%
\stepcounter{equation}\edef\@tempa{\theequation}%
\let\c@equation\c@subeqncnt\c@subeqncnt\z@
\edef\theequation{\@tempa\noexpand\thesubeqncnt}}

\makeatother

\begin{document}

\titlepage

\title{The Static Elliptic $N$-soliton Solutions of the KdV Equation} 
\author{Masahito Hayashi\thanks{masahito.hayashi@oit.ac.jp}\\
Osaka Institute of Technology, Osaka 535-8585, Japan\\
Kazuyasu Shigemoto\thanks{shigemot@tezukayama-u.ac.jp} \\
Tezukayama University, Nara 631-8501, Japan\\
Takuya Tsukioka\thanks{tsukioka@bukkyo-u.ac.jp}\\
Bukkyo University, Kyoto 603-8301, Japan\\
}
\date{\empty}


\maketitle
\abstract{
Regarding $N$-soliton solutions, 
the trigonometric type, the hyperbolic type, and the exponential type
solutions have been well studied. 
While for the elliptic type solution, we know only the one-soliton 
solution so far. 
Using the commutative B\"{a}cklund transformation, we have succeeded in
constructing the KdV static elliptic $N$-soliton solution, which means that 
we have obtained infinitely many solutions for the $\wp$-function type 
differential equation.
}

\section{Introduction} 

\setcounter{equation}{0}

Quite interesting nonperturbative phenomena are discovered 
by studies of soliton systems. 
Since 
the inverse scattering 
method~\cite{Gardner,Lax,Zakhrov}, 
many interesting developments 
have been done 
including the AKNS formulation~\cite{Ablowitz}, 
the B\"{a}cklund
transformation~\cite{Wahlquist,Wadati1,Wadati2}, 
the Hirota equation~\cite{Hirota1,Hirota2}, 
the Sato theory~\cite{Sato}, 
the vertex construction of the soliton solution~\cite{Date}, 
and the Schwarzian type mKdV/KdV equation~\cite{Weiss}. 
Our understanding of the soliton has been 
still in progress. 

The name ``soliton'' has come through studies of the KdV equation. 
In nontrivial solutions of the KdV equation, there exists a 
solitary wave solution which can be regarded as an excitation of 
particle i.e.\ soliton.    
The KdV equation can also provide interacted configurations by its solutions.  
With proper time dependence, collision phenomena of solitons 
can be captured by such multi soliton solutions. 
A soliton solution can be visualized as a spatially localized object,  
and we in this paper refer this definition for soliton solutions.    
Having $N$ localized excitation, we call this as an ``$N$-soliton''
solution. 

Since the KdV equation is a nonlinear differential
equation, 
it has been not obvious to find out $N$-soliton solutions due to lack of 
linear superposition. 
Nevertheless, it is now to be standard to construct $N$-soliton solutions from 
one soliton solutions
by the B\"{a}cklund transformation.
In other wards, we could consider such a nontrivial 
nonlinear superposition in special cases.   

In order to solve nonlinear differential equations, 
underlying symmetries which the systems possess 
may play a crucial role. 
In the AKNS formulation, the soliton equations such as the 
KdV, the mKdV, and the sine-Gordon equations are obtained as 
the integrability condition of real $2\times 2$ matrix, 
which means the symmetry of the soliton systems 
lies on 
the M\"{o}bius (GL(2,$\mathbb{R}$)) group symmetry. 

In our previous paper~\cite{Hayashi},  we have studied the algebraic 
construction of the $N$-soliton solutions. 
Using pieces of one-soliton solutions 
obtained by directly solving differential equations, 
we have algebraically constructed $N$-soliton solutions by using 
the commutative 
B\"{a}cklund transformation for the KdV, the mKdV, and the sine-Gordon 
equations. 
In this algebraic construction, the commutative 
subgroup, i.e.\ commutative B\"{a}cklund transformation of 
the M\"{o}bius group symmetry, has been essential. 
The $N$-soliton solutions which we had obtained were in the 
hyperbolic type (the exponential type).  
The addition formula of the hyperbolic  
function  such as $\tanh(x+\xi)$ gives 
$$
\tanh(x+\xi)=\frac{\alpha \tanh x+\beta}
{\gamma \tanh x+\delta} , 
$$
which is the global M\"{o}bius transformation 
with $\alpha=1$, $\beta=\tanh \xi$ , $\gamma=\tanh \xi$, and 
$\delta=1$. 
The algebraic $N$-soliton construction in the previous 
paper~\cite{Hayashi} 
is the result from the local commutative M\"{o}bius transformation. 
This could be a realization of nontrivial superposition. 

So far we know only one-soliton solution of the elliptic type. 
Considering the Ising model, 
we observe that 
the SU(2) group symmetry and the elliptic function appear and they 
are mutually connected~\cite{Onsager,Baxter}.  
As the structures of the SU(2) and  GL(2,$\mathbb{R}$) is similar, 
we suppose it may be possible to access to elliptic $N$-soliton solutions
through the commutative B\"{a}cklund transformations.  

The paper is organized as follows: 
In section $2$, we briefly review the previous studies and make some 
preparations. 
Then explicit constructions of the static elliptic $N$-soliton 
solutions are presented in section $3$. 
We devote the final section to the summary and the discussions.   


\section{The KdV One-Soliton Solutions} 

\setcounter{equation}{0}

\subsection{The KdV equation and its elliptic one-soliton solution}

The KdV equation is given by~\footnote{Indices $x$ in 
expressions $u_x, u_{xx}, \cdots$ imply the 
partial derivative with respect to $x$.  
We use this notation throughout the paper.} 
\begin{equation}
  u_t-u_{xxx}+6 u u_x=0. 
\label{2e1}
\end{equation}
In order to find the one-soliton solution,
we assume a linear dependence for $x$ and $t$ as 
$ax+bt+\delta=:X$ with constant parameters $a$, $b$, and $\delta$. 
Setting $u(x,t)=2 U(X)$ with the variable $X$,  
the KdV equation (\ref{2e1}) becomes 
$$
 bU_X-a^3U_{XXX}+12aUU_X=
 -a^3U_{XXX}+12 a\Big(U+\frac{b}{12a}\Big)U_X=0.
$$
Redefining $\widehat{U}=U+\dfrac{b}{12a}$, we arrive at 
\begin{equation}
\widehat{U}_{XXX} =\frac{12}{a^2} \widehat{U}\widehat{U}_X.
 \label{2e3}
\end{equation}

Now let us remind ourselves 
the Weierstrass's $\wp$-function which satisfies
\begin{subequations}
\begin{align}
\wp_x(x)^2&=4 \wp(x)^3-g_2\wp(x)-g_3 
\nonumber \\ 
&=4 (\wp(x)-e_1) (\wp(x)-e_2) (\wp(x)-e_3), 
\label{2e4a} \\
\wp_{xx}(x)&=6 \wp(x)^2-\frac{g_2}{2}, 
\label{2e4b}\\
\wp_{xxx}(x)&=12 \wp(x)\wp_{x}(x), 
\label{2e4c} 
\end{align}
\end{subequations}

\vspace*{-4mm}
\noindent
where $e_1$, $e_2$, and $e_3$ points are determined 
through usual Vieta's root formulas: 
\begin{equation}
e_1+e_2+e_3=0, \quad 
e_1e_2+e_2e_3+e_3e_1=-\dfrac{1}{4}g_2, \quad 
e_1e_2e_3=\dfrac{1}{4}g_3. 
\end{equation}
Eqs.(\ref{2e4b}) and (\ref{2e4c}) are directly derived from Eq.(\ref{2e4a}).  

Thanks to Eq.(\ref{2e4c}), it is easy to observe that 
the $\wp$-function is 
a solution of the KdV equation (\ref{2e3}) 
with $\widehat{U}(X)=a^2 \wp(X)$. 
Thus, in the original form, 
we have the elliptic one-soliton solution
\begin{equation}
u(x,t)=2a^2 \wp(ax+bt+\delta)-\frac{b}{6a}.
\label{2e5}
\end{equation}
We discuss the time-dependent $N$-soliton solution
 in the summary and discussions, so that we first construct the static $N$-soliton 
solutions.
Thus, we  concentrate on the static case hereafter.  
The static elliptic one-soliton solution now has the form from 
Eq.(\ref{2e5}), 
\begin{equation}
u(x)=2a^2 \wp(ax+\delta).
\label{2e6}
\end{equation}

Before closing this subsection, 
it should be mentioned that the KdV equation 
can be rewritten as the $\wp$-function type differential equation. 
Integrating the static version of the KdV equation (\ref{2e1})
twice,  
we directly obtain 
\begin{equation}
 u_{x}(x)^2=2 u(x)^3+Cu(x)+D,
\label{2e6+} 
\end{equation}
with integration constants $C$ and $D$. 
Sending the constants to $C=-2g_2$ and $D=-4g_3$, 
and redefining the function as $u(x)=2h(x)$, 
it is easy to see that Eq.(\ref{2e6+}) turns to be 
the same form as Eq.(\ref{2e4a}), 
\begin{equation}
 h_x(x)^2=4 h(x) ^3-g_2 h(x) -g_3. 
\label{3e9b}
\end{equation}

\subsection{Another static elliptic one-soliton solution}

The Jacobi's elliptic function ${\rm sn}(x)$ satisfies 
the following differential equation: 
\begin{equation}
 f_x(x)^2=(1-f(x)^2)(1-k^2 f(x)^2) ,
\label{2e7}
\end{equation}
with $k^2=(e_2-e_3)/(e_1-e_3)$.
For any functions $f(x)$ which satisfy Eq.(\ref{2e7}), the following function $h(w)$
\begin{equation}
h(w)=h\Big(\dfrac{x}{\sqrt{e_1-e_3}}\Big)
=e_3+\frac{e_1-e_3}{f^2(x)}, 
\label{2e8}
\end{equation}
obeys the $\wp$-function type differential equation
\begin{equation}
h_w(w)^2=4 h(w) ^3-g_2 h(w) -g_3. 
\label{2e9}
\end{equation}
It is easy to show that $f(x)=1/(k\, {\rm sn}(x))$ also satisfies
Eq.(\ref{2e7}). Then we find the following function $h_1(w)$  
\begin{equation}
 h_1(w)=h_1 \Big(\dfrac{x}{\sqrt{e_1-e_3}}\Big)=
e_3+ (e_1-e_3) k^2 {\rm sn}^2(x), \quad
\label{2e10}
\end{equation}
satisfies the $\wp$-function type differential equation (\ref{2e9}).
Since the $\wp(w)$ function and the ${\rm sn}(x)$ function are connected
in the form 
\begin{equation}
\wp(w)=\wp\Big(\dfrac{x}{\sqrt{e_1-e_3}}\Big)
=e_3+\frac{e_1-e_3}{{\rm sn}^2(x)}, 
\label{2e11}
\end{equation}
the function $h_1(w)$ defined by Eq.(\ref{2e10}) becomes 
the M\"{o}bius transformed form of the $\wp(w)$ function
\begin{equation}
 h_1(w)=\frac{\alpha\wp(w)+\beta}{\gamma\wp(w)+\delta},
\label{2e12}
\end{equation}
with $\alpha \delta-\beta \gamma \ne 0$, 
and $\alpha=-e_1-e_2$, $\beta={e_1}^2+{e_2}^2+3 e_1 e_2$, 
$\gamma=1$, $\delta=e_1+e_2$.  
Then we get another static elliptic one-soliton solution 
$u(w)=2h_1(w)$. 

\subsection{Hyperbolic one-soliton solution 
by the B\"{a}cklund transformation}

Let us now introduce the B\"{a}cklund transformation which can 
generate $N$-soliton solutions. 
Using the variable $z_x(x)=u(x)$, the B\"{a}cklund transformation 
of the KdV equation~\cite{Wahlquist} is given by 
\begin{equation}
   z'_x(x)+z_x(x)=-\frac{\lambda^2}{2}+\dfrac{(z'(x)-z(x))^2}{2}, 
\label{2e13}
\end{equation}
with new arbitrary parameter $\lambda$. 
For the given soliton solution $z(x)$,  Eq.(\ref{2e13})
provides a condition  that the new soliton solution $z'(x)$ 
must satisfy. 
It should be noted that 
this B\"{a}cklund transformation is the only commutative one, as far as
we know. 

In our previous paper~\cite{Hayashi}, 
we have constructed $N$-soliton solutions of the mKdV equation
by using the KdV-type B\"{a}cklund transformation~\cite{Wahlquist} 
instead of the mKdV-type  B\"{a}cklund transformation~\cite{Wadati1} 
by making the 
connection between the mKdV equation and the KdV equation through the 
Miura transformation. 
The reason why we can construct $N$-soliton solutions by 
the KdV-type B\"{a}cklund transformation is that 
it is the only 
commutative  one. 
We had emphasized in our previous paper~\cite{Hayashi} 
that commutative B\"{a}cklund transformations play an important role 
to construct $N$-soliton solutions algebraically. 
 
Let us make use of the B\"acklund transformation to obtain 
soliton solution.
As the trivial solution, we have $z(x)=0$.
In this case, 
the B\"{a}cklund transformation Eq.(\ref{2e13}) tells us that 
another soliton solution $z'(x)$ satisfies the following 
``differential equation'',   
\begin{equation*}
 z'_x=\frac{1}{2}(z'^2-\lambda^2).  
\label{2e14}
\end{equation*}
One can solve the differential equation and get 
the hyperbolic type solution,
\begin{equation*}
z'=-\lambda\tanh\Big(\frac{\lambda x+\delta}{2}\Big), 
\label{2e15}
\end{equation*}
with an arbitrary parameter $\delta$. 
Thus,  if we put $z(x)=0$, we cannot obtain the elliptic $N$-soliton solution 
via B\"{a}cklund transformation.
In the next section, we will show that both $z(x)$ and $z'(x)$ can be
non-zero in the B\"{a}cklund transformation Eq.(\ref{2e13}). 
We can take elliptic type functions in such a way as both solutions are 
consistent with the KdV-type 
B\"{a}cklund transformation Eq.(\ref{2e13}).  
This fact is the key point for our construction 
of the elliptic $N$-soliton solutions. 

\section{The Static Elliptic $\bm N$-soliton Solutions}
\setcounter{equation}{0}

We work with the B\"{a}cklund transformation of the KdV equation 
given by Eq.(\ref{2e13}). 

We prepare two elliptic one-soliton solutions 
which have the forms of Eq.(\ref{2e6}), 
\begin{align}
u(x)&=2{a_1}^2 \wp(a_1x+\delta_1)=: z_x(x), 
\label{3e1}\\
u'(x)&=2{a_2}^2 \wp(a_2x+\delta_2)=: z'_x(x), 
\label{3e2}
\end{align}
where we have introduced $z_x(x)$ and $z'_x(x)$ 
for the sake of using B\"{a}cklund transformation. 

Using the relation between the $\wp$- and $\zeta$-functions, 
\begin{equation}
\zeta_x(x)
=-\wp(x), 
\end{equation}
we have 
\begin{align}
z(x)&=-2a_1\zeta(a_1 x+\delta_1)+\eta_1,
\label{3e3}\\
z'(x)&=-2a_2 \zeta(a_2 x+\delta_2)+\eta_2, 
\label{3e4}
\end{align}
with integration constants $\eta_1$ and $\eta_2$. 
Then we examine whether we can arrange these $z(x)$, $z'(x)$ 
to satisfy the B\"{a}cklund transformation Eq.(\ref{2e13}). 
Substituting Eqs.(\ref{3e3}) and (\ref{3e4}) 
into Eq.(\ref{2e13}), we have 
\begin{align}
&2{a_2}^2 \wp(a_2x+\delta_2)+2{a_1}^2 \wp(a_1x+\delta_1) \nonumber\\
&
=-\frac{\lambda^2}{2}
+\frac{1}{2}
\Big(-2 a_2 \zeta (a_2 x+\delta_2)+\eta_2+2 a_1 \zeta (a_1 x+\delta_1)
-\eta_1\Big)^2.
\label{3e5}
\end{align}
We now look at the relation, 
\begin{equation}
\wp(u+v)+\wp(u)+\wp(v)=\big(\zeta(u+v)-\zeta(u)-\zeta(v)\big)^2,  
\label{3e6}
\end{equation}
and adjust the parameters in (\ref{3e3}) and (\ref{3e4}) 
so as to get consistency between Eqs.(\ref{3e5}) and (\ref{3e6}).  
We first take $a_1=a_2$ and put $\eta_1=0$ without loss of generality
by the constant shift of $x$. 
Thus, choosing the parameters as 
\begin{equation*}
a_1=a_2=1, \quad 
\delta_1=0, \quad \delta_2=\delta, \quad   
\eta_1=0, \quad 
\eta_2=2\zeta(\delta), \quad  \lambda^2/4=\wp(\delta), 
\end{equation*}
we can accommodate Eq.(\ref{3e5}) to the following form
\begin{equation}
\wp(x+\delta)+\wp(x)+\wp(\delta)=
\big(\zeta(x+\delta)
-\zeta(x)-\zeta(\delta)\big)^2, 
\label{3e7}
\end{equation}
which suits the relation Eq.(\ref{3e6}). 
As the result, we can obtain the pair of elliptic one-soliton solutions 
$z(x)$ and $z'(x)$ in the B\"{a}cklund transformation Eq.(\ref{2e13}),
which are consistently coexist, in the form 
$z(x)=-2\zeta(x)$ and $z'(x)=-2\big(\zeta(x+\delta)-\zeta(\delta)\big)$.  
By changing 
the parameter $\delta$, 
we obtain infinitely many one-soliton solutions:
\begin{align}
z&=-2\zeta(x)=:z_0,
\label{3e8}\\
z'&=-2(\zeta(x+\delta_i)-\zeta(\delta_i))=:z_i.
\label{3e9}
\end{align}

In the next section, 
using these one-soliton solutions $z_0(x)$ and $z_i(x)$, we can algebraically 
construct $N$-soliton solutions by the commutative B\"{a}cklund 
transformation. 
In terms of
$z_x(x)=u(x)=2h(x)$, 
we fix our ``KdV equation'' to be solved as 
Eq.(\ref{2e6+}) with $C=-2g_2$ and $D=-4g_3$, i.e.\, 
\begin{equation}
{z_{xx}}^2=2 {z_x}^3-2g_2z_x-4g_3, 
\label{3e9a}
\end{equation}
which can be related with the $\wp$-function type differential
equation (\ref{3e9b}).  

\subsection{The static elliptic $\bm{(2+1)}$-soliton solution}

Using three elliptic one-soliton solutions given in Eqs.(\ref{3e8}) and
(\ref{3e9}), i.e., 
\begin{align*}
z_0&=-2\zeta(x), 
\\
z_1&=-2(\zeta(x+\delta_1)-\zeta(\delta_1)), 
\\ 
z_2&=-2(\zeta(x+\delta_2)-\zeta(\delta_2)),
\end{align*}
we will algebraically construct an $N$-soliton solution 
by the B\"{a}cklund  transformation.

We prepare the B\"{a}cklund  transformations (\ref{2e13}) 
which provide $z_1(x)$ and $z_2(x)$ from $z_0(x)$ separately,  
\begin{subequations}
\begin{align}
z_{1, x}+z_{0, x}
&=-\dfrac{{\lambda_1}^2}{2}+\dfrac{(z_1-z_0)^2}{2},
\label{3e10a}\\
z_{2, x}+z_{0, x}
&=-\dfrac{{\lambda_2}^2}{2}+\dfrac{(z_2-z_0)^2}{2},
\label{3e10b}
\end{align}
with $\lambda^2_1=4\wp(\delta_1)$,  $\lambda^2_2=4\wp(\delta_2)$. 
We then assume the commutativity to access to $z_{12}(x)$ 
via $z_1(x)$ and $z_2(x)$, 
\begin{align}
 z_{12, x}+z_{1, x}
&=-\dfrac{{\lambda_2}^2}{2}+\dfrac{(z_{12}-z_1)^2}{2},
\label{3e10c}\\
z_{12, x}+z_{2, x}
&=-\dfrac{{\lambda_1}^2}{2}+\dfrac{(z_{12}-z_2)^2}{2}.
\label{3e10d}
\end{align}
\end{subequations}
Schematically, the commutativity is displayed as the following diagram: 
$$
\xymatrix@ur{
z_0 \ar[r] \ar[d] & z_1 \ar[d] \\
z_2 \ar[r] & z_{12} & 
}
$$

Manipulating 
``Eq.(\ref{3e10a})$-$Eq.(\ref{3e10b})$-$Eq.(\ref{3e10c})$+$Eq.(\ref{3e10d})'',  
we can excavate the relation  
\begin{align}
z_{12}
&=
z_0+\frac{{\lambda_1}^2-{\lambda_2}^2}{z_1-z_2}
\nonumber \\
&=-2\zeta(x)
-\dfrac{2(\wp(\delta_1)-\wp(\delta_2))}
{\zeta(x-\delta_1)-\zeta(x-\delta_2)-\zeta(\delta_1)+\zeta(\delta_2)}.
\label{3e11}
\end{align}
We can check that 
Eq.(\ref{3e11}) is consistent with 
the series of Eqs.(\ref{3e10a})-(\ref{3e10d}), so that our 
assumption of the commutativity is guaranteed.  
We have also confirmed numerically by Mathematica 
that our solution $z_{12}(x)$ really satisfies Eq.(\ref{3e9a}). 
Therefore, the function $z_{12}(x)$ which is given by Eq.(\ref{3e11}) 
is the new soliton solution of the
static KdV equation Eq.(\ref{3e9a}). 

In the solution, $z_1(x)$ and $z_2(x)$ 
come in the cyclic symmetric form, but $z_0(x)$, $z_1(x)$ and $z_2(x)$ do
not, so that we call this solution 
as the static elliptic $(2+1)$-soliton solution. 

\begin{figure}[h!]
\qquad\quad
 \begin{minipage}{0.4\hsize}
  \begin{center}
\hspace{-10.1mm}
   \includegraphics[width=80mm]{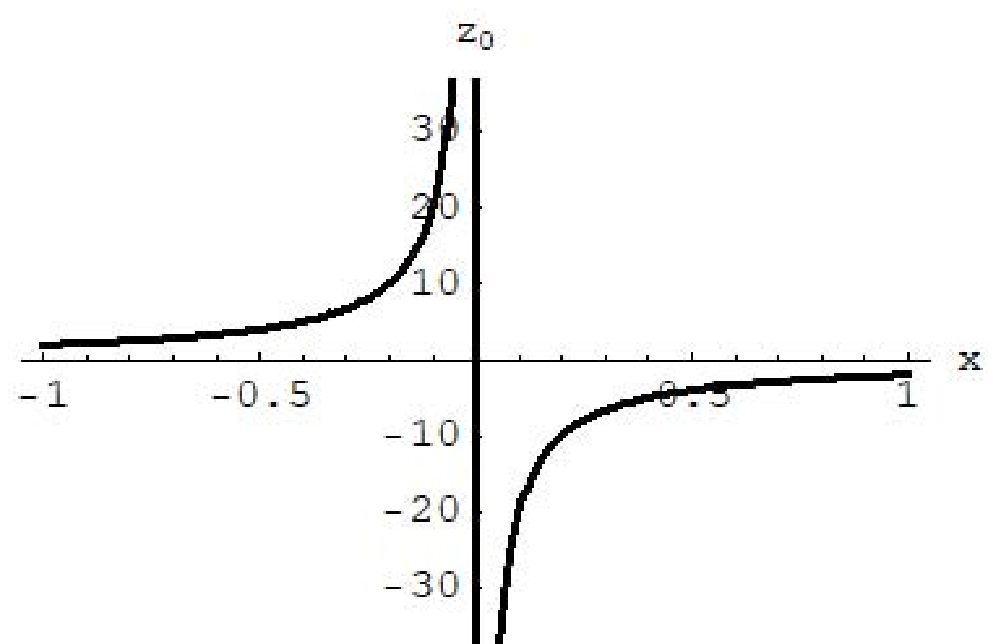}
  \end{center}
 \caption{
$z_0(x)=-2\zeta(x)$ \, 
\protect\newline
\hspace*{17mm} with $g_2=0.3$, \, $g_3=0.7$
}
  \label{fig1}
 \end{minipage}
 \qquad
 \lower1.3ex\hbox{%
 \begin{minipage}{0.4\hsize}
  \begin{center}
    \includegraphics[width=80mm]{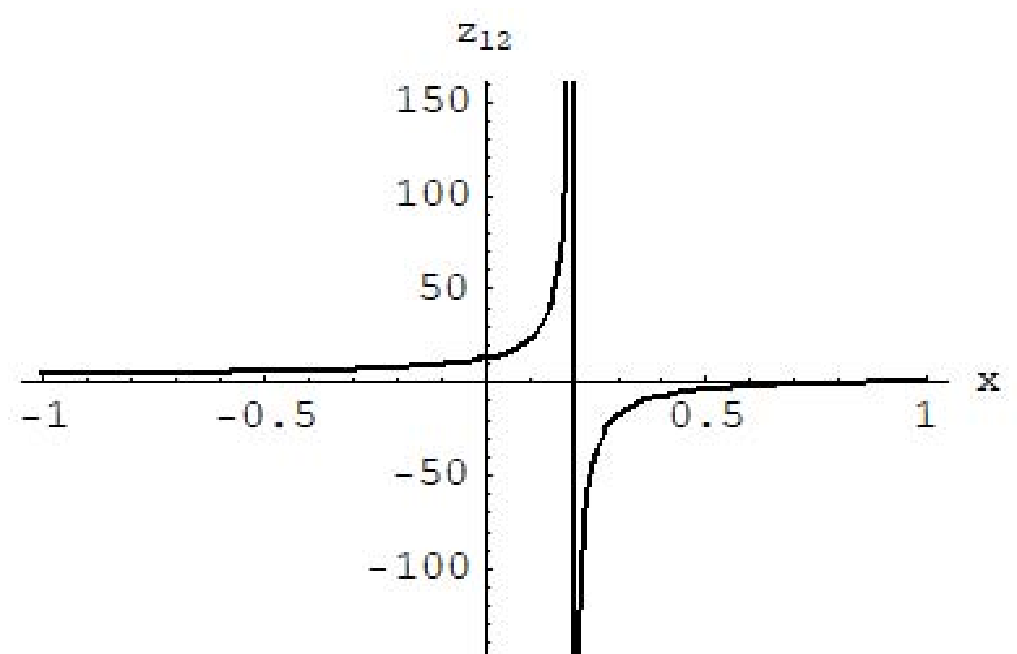}
  \end{center}
  \caption{$z_{12}(x)$ 
\protect\newline 
\hspace*{17mm} with $g_2=0.3$, \, 
$g_3=0.7$, 
\protect\newline 
\hspace*{26.5mm} $\delta_1=-0.02$, \, 
$\delta_2=0.04$}
  \label{fig2}
 \end{minipage}
 }
\end{figure}

We sketch the graphs of $z_0(x)$ and $z_{12}(x)$ in Figure \ref{fig1} and 
Figure \ref{fig2}, respectively. 
We can observe that the pole at $x=0$ in $z_0(x)$ disappears in $z_{12}(x)$, 
which can be seen by expanding  Eq.(\ref{3e11}) around $x=0$. 
We can also see that $z_{12}(x)$ becomes narrower than $z_0(x)$ in width. 

Taking the derivative of Eq.(\ref{3e11}), we have
\begin{align}
u&=z_{12, x}
\nonumber \\
&= 2\wp(x)-\frac{2(\wp(\delta_1)-\wp(\delta_2))
\big(\wp(x+\delta_1)-\wp(x+\delta_2)\big)}
{\big(\zeta(x+\delta_1)-\zeta(x+\delta_2)-\zeta(\delta_1)
+\zeta(\delta_2)\big)^2},
\label{3e11a}
\end{align}
which corresponds to the static elliptic KdV $(2+1)$-soliton solution 
for Eq.(\ref{2e1}).

\subsection{The static elliptic $\bm 3$-soliton solution} 

Let us construct another type of an $N$-soliton solution. 
In addition to the previous solutions $z_1(x)$, $z_2(x)$, and $z_{12}(x)$ 
given from $z_0(x)$,   
we here prepare $z_{13}(x)$ and $z_3(x)$ which are also constructed 
from $z_0(x)$. 
Thus, we have additional relations
\begin{subequations}
\begin{align}
z_{3, x}+z_{0, x}&=-\frac{{\lambda_3}^2}{2}+\frac{(z_{3}-z_0)^2}{2},
\label{3e12a}\\
z_{13}&=z_0+\frac{{\lambda_1}^2-{\lambda_3}^2}{z_1-z_3}.
\label{3e12b}
\end{align}
\end{subequations}

\vspace*{-4mm}
\noindent
Using the B\"{a}cklund  transformations 
and here assuming the following commutativity 
$$
\xymatrix@ur{
z_1 \ar[r] \ar[d] & z_{12} \ar[d] \\
z_{13} \ar[r] & z_{123} & 
}
$$
we have 
\begin{subequations}
\begin{align}
z_{12, x}+z_{1, x}&=-\frac{{\lambda_2}^2}{2}+\frac{(z_{12}-z_1)^2}{2},
\label{3e14a}\\
z_{13, x}+z_{1, x}&=-\frac{{\lambda_3}^2}{2}+\frac{(z_{13}-z_1)^2}{2},
\label{3e14b}\\
z_{123, x}+z_{12, x}&=-\frac{{\lambda_3}^2}{2}+\frac{(z_{123}-z_{12})^2}{2},
\label{3e14c}\\
z_{123, x}+z_{13, x}&=-\frac{{\lambda_2}^2}{2}+\frac{(z_{123}-z_{13})^2}{2}.
\label{3e14d}
\end{align}
\end{subequations}

Considering ``Eq.(\ref{3e14a})$-$Eq.(\ref{3e14b})$-
$Eq.(\ref{3e14c})$+$Eq.(\ref{3e14d})'', we obtain the following relations  
\begin{align}
z_{123}
&=z_1+\frac{{\lambda_2}^2-{\lambda_3}^2}{z_{12}-z_{13}}  \nonumber\\
   &=-\frac{({\lambda_1}^2-{\lambda_2}^2)z_1 z_2
+({\lambda_2}^2-{\lambda_3}^2)z_2 z_3+({\lambda_3}^2-{\lambda_1}^2)z_3 z_1}
      {({\lambda_1}^2-{\lambda_2}^2)z_3+({\lambda_2}^2-{\lambda_3}^2)z_1
+{(\lambda_3}^2-{\lambda_1}^2)z_2}, 
\label{3e17}
\end{align}
with 
\vspace*{-3mm}
\begin{align*}
z_0&=-2\zeta(x), & z_1&=-2(\zeta(x+\delta_1)-\zeta(\delta_1)), 
\\ 
z_2&=-2(\zeta(x+\delta_2)-\zeta(\delta_2)), &  
z_3&=-2(\zeta(x+\delta_3)-\zeta(\delta_3)), 
\end{align*}
and
\vspace*{-3mm}
\begin{align*}
{\lambda_1}^2&=4\wp(\delta_1), 
& {\lambda_2}^2&=4\wp(\delta_2),  &{\lambda_3}^2&=4\wp(\delta_3). 
\end{align*}


\noindent
We have checked that 
Eq.(\ref{3e17}) is consistent with 
the series of Eqs.(\ref{3e14a})-(\ref{3e14d}) and our 
assumption of the commutativity is guaranteed.  
We have also confirmed numerically by Mathematica 
that our solution $z_{123}(x)$ really satisfies Eq.(\ref{3e9a}).
Therefore, 
the function $z_{123}(x)$  is
the new soliton solution
of the static KdV equation (\ref{3e9a}).  

Because of the commutativity of the B\"{a}cklund transformation, 
the expression in Eq.(\ref{3e17}) becomes in the cyclic symmetric 
form for $z_1(x)$, $z_2(x)$ and $z_3(x)$, 
which confirms that  $3!$-independent construction of $z_{123}(x)$ 
gives the same result as above. 
Then we call this solution as the static elliptic 3-soliton solution.  

We can recursively show the commutativity 
of the B\"{a}cklund transformation by  identifying
$$
z_{1,2,\cdots,i-1}\rightarrow z_0, \qquad
z_{1,2,\cdots,i-1,i}\rightarrow z_1, \qquad 
z_{1,2,\cdots,i-1,i+1}\rightarrow z_2, \qquad
z_{1,2,\cdots,i-1,i,i+1}\rightarrow z_{12}, 
$$ 
in the proof 
of $2+1$-soliton solution.

\begin{figure}[h!]
  \begin{center}
   \includegraphics[width=80mm]{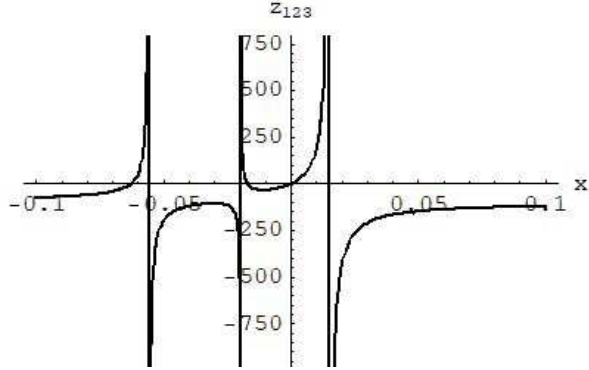}
  \end{center}
  \caption{$z_{123}(x)$ with $g_2=0.3$, 
$g_3=0.7$, $\delta_1=-0.02$, $\delta_2=0.03$, $\delta_3=0.05$}
  \label{fig3}
\end{figure}

We sketch the graph of $z_{123}(x)$ in Figure \ref{fig3}. 
We can see three localized clusters in this solution.

%
\subsection{The static elliptic (4+1)-soliton solution and 5-soliton
  solution}

We can further proceed to construct the static elliptic solutions. 
The KdV $(4+1)$-soliton solution for 
(\ref{3e9a}) can be obtained as 
\begin{equation}
z_{1234}
=z_{12}+\frac{{\lambda_3}^2-{\lambda_4}^2}{z_{123}-z_{124}}, 
\label{3e18}
\end{equation}
with Eq.(\ref{3e17}) and its cyclic symmetric expression 
$z_{124}(x)$ and ${\lambda_i}^2=4 \wp(\delta_i)$. 
The expression
of $z_{1234}(x)$ with $z_0(x)$, $z_1(x)$, $z_2(x)$, $z_3(x)$, and $z_4(x)$ is 
given explicitly in the form
\begin{equation}
z_{1234}=z_0+\frac{G_{1234}}{F_{1234}},  
\label{3e19}
\end{equation}
with
\begin{align*}
F_{1234}&=\sum^4_{i,j,k,l=1} \epsilon^{ijkl}
 ({\lambda_i}^2-{\lambda_j}^2)({\lambda_k}^2-{\lambda_l}^2)  z_i z_j, 
\\
G_{1234}&=-2 \sum^4_{i,j,k,l=1}  \epsilon^{ijkl}
 {\lambda_i}^2 {\lambda_j}^2({\lambda_i}^2-{\lambda_j}^2) z_k, 
\end{align*}
where $\epsilon^{ijkl}$ is the totally antisymmetric tensor with 
$\epsilon^{1234}=1$.

The static elliptic KdV $5$-soliton solution for Eq.(\ref{3e9a}) 
is given by 
\begin{equation}
z_{12345}
=z_{123}+\frac{{\lambda_4}^2-{\lambda_5}^2}{z_{1234}-z_{1235}}.
\label{3e22}
\end{equation}
The expression of $z_{12345}(x)$ 
with $z_0(x)$, $z_1(x)$, $z_2(x)$, $z_3(x)$, $z_4(x)$, and $z_5(x)$ 
is given in the form
\begin{equation}
z_{12345}=\frac{G_{12345}}{F_{12345}},  
\label{3e23}
\end{equation}
with
\begin{align*}
F_{12345} 
&=\sum^5_{i,j,k,l,m=1} \epsilon^{ijklm}
 ({\lambda_i}^2-{\lambda_j}^2)({\lambda_k}^2-{\lambda_l}^2)
({\lambda_l}^2-{\lambda_m}^2)({\lambda_m}^2-{\lambda_k}^2) z_i z_j,
\\
G_{12345}
&=\sum^5_{i,j,k,l,m=1} \epsilon^{ijklm}
 ({\lambda_i}^2-{\lambda_j}^2)({\lambda_k}^2-{\lambda_l}^2)
({\lambda_l}^2-{\lambda_m}^2)({\lambda_m}^2-{\lambda_k}^2) z_k z_l z_m. 
\end{align*}

We have numerically confirmed that both the static elliptic 
$(4+1)$-soliton and the $5$-soliton solutions 
really satisfy the 
static KdV equation (\ref{3e9a}). 

In the same manner, we could recursively construct (1+(even number))-soliton 
solutions and (odd number)-soliton solutions. 
In the (odd number)-soliton solutions, 
$z_0$ cancels out and does not appear in the final soliton solutions. 
General structures of static elliptic solutions could be discussed elsewhere.

\section{Summary and Discussions} 
\setcounter{equation}{0}

Regarding soliton solutions for the elliptic type, only the one-soliton
solution has been available so far.   
We have obtained the KdV static elliptic $N$-soliton solutions
by using the commutative B\"{a}cklund transformations.  
We understand that the key point of the algebraic construction of 
the KdV static elliptic $N$-soliton solution is the existence of the
M\"{o}bius (GL(2,$\mathbb{R}$)) group symmetry and the 
one-soliton solutions of the algebraic functions such as the  
trigonometric, the hyperbolic or the elliptic types for the 
KdV equation.  
The local algebraic addition formula of the algebraic functions,  
which comes from the commutative B\"{a}cklund transformation,  
seems to be essential. 

For the time-dependent solution, we can construct a certain 
time-dependent solution by the static solution, which can be constructed 
in our paper, by just the following replacement. We denote the static solution
$u^{({\rm static})}(x)$,  which can be written in the form 
$$
u^{({\rm static})}(x)=F\big(f_1(x+\delta_1),\, f_2(x+\delta_2), \, \cdots\big) .
$$
Then we replace $x\rightarrow x+b t$ in this static solution and we have 
$$
u^{({\rm static})}(x+b t)=F\big(f_1(x+b t+\delta_1),\, 
f_2(x+b t+\delta_2), \, \cdots\big) .
$$ 
Through the following manipulation, 
\begin{align*}
u^{({\rm static})}_t(x+b t)
&=F_t\big(f_1(x+b t+\delta_1),\, f_2(x+b t+\delta_2), \, \cdots\big) 
\\
&=b F_x\big(
f_1(x+bt+\delta_1),\, f_2(x+bt+\delta_2), \, \cdots\big)
\\
&=b u^{({\rm static})}_x(x+bt), 
\end{align*}
we find 
$$
\widehat{u}(x,t)=u^{({\rm static})}(x+bt)-\dfrac{b}{6}
$$ 
becomes the 
time-dependent solution of the KdV equation 
$$ 
\widehat{u}_t(x,t)-\widehat{u}_{xxx}(x,t)
+6 \widehat{u}(x,t) \widehat{u}_x(x,t)=0, 
$$
by using
$$
u^{({\rm static})}_t(x+b t)= b u^{({\rm static})}_x(x+bt), 
\quad  
u^{({\rm static})}_{xxx}(x+b t)
=6u^{({\rm static})}(x+b t) u^{({\rm static})}_x(x+b t).
$$
This time-dependent solution $\widehat{u}(x,t)$ is the special generalization 
of the time-dependent elliptic solution Eq.(\ref{2e5}).

Our $N$-solitons and well-known one-soliton as elliptic type 
solutions of the KdV equation are both singular.
Originally the KdV equation is derived as the wave equation of the 
shallow water by taking the special limit. Then
the KdV equation is an idealistic equation, so that  
our singular solutions will correspond to the 
much milder solitary waves in the real shallow water.
However, what we prefer here is to emphasize a deep relationship  
between mathematics and underlying physics. 
If we consider the $\wp$-function type differential equation as the 
static KdV equation,  we have infinitely many elliptic soliton
solutions for the $\wp$-function type differential equation. 
In other wards, we find a family of the $\wp$-function via 
the physical integrable KdV system. 
This might be quite interesting not only for physics but also for mathematics.


\end{document}